\documentclass[11pt]{article}
\usepackage{times}
\usepackage{amsmath, amssymb, amsthm}
\usepackage{url}
\usepackage{graphicx}
\usepackage{hyperref}
\usepackage{cleveref}

\title{(G)I-DLE: Generative Inference via Distribution-preserving Logit Exclusion with KL Divergence Minimization for Constrained Decoding}
\author{Hanwool Lee \\
Shinhan Securities Co. \\
\texttt{gksdnf424@gmail.com}}

\begin{document}
\maketitle

\begin{abstract}
We propose (G)I-DLE, a new approach to constrained decoding that leverages KL divergence minimization to preserve the intrinsic conditional probability distribution of autoregressive language models while excluding undesirable tokens. Unlike conventional methods that naively set banned tokens’ logits to $-\infty$, which can distort the conversion from raw logits to posterior probabilities and increase output variance, (G)I-DLE re-normalizes the allowed token probabilities to minimize such distortion. We validate our method on the K\(^2\)-Eval dataset, specifically designed to assess Korean language fluency, logical reasoning, and cultural appropriateness. Experimental results on Qwen2.5 models (ranging from 1.5B to 14B) demonstrate that (G)I-DLE not only boosts mean evaluation scores but also substantially reduces the variance of output quality.
\end{abstract}

\section{Introduction}
Constrained decoding plays a crucial role in neural text generation by ensuring that models adhere to predefined content restrictions. Traditional methods typically enforce such constraints by setting the logits of banned tokens to $-\infty$. Although effective for strict exclusion, these approaches can inadvertently distort the underlying probability distribution, particularly when the banned tokens hold a significant probability mass in multilingual models. This issue is especially problematic in domain-specific settings, such as generating culturally appropriate Korean text. 

In this work, we introduce (G)I-DLE (Generative Inference via Distribution-preserving Logit Exclusion), a method that formulates constrained decoding as a KL divergence minimization problem. By re-normalizing the allowed token probabilities, (G)I-DLE minimizes the distortion induced by naive masking. We evaluate our method on the K\(^2\)-Eval dataset, which consists of 90 open-ended questions designed to test Korean language fluency, logical reasoning, and cultural knowledge. Experiments on Qwen2.5 models of various sizes demonstrate that (G)I-DLE yields more fluent, coherent, and culturally appropriate responses with reduced output variance compared to both baseline generation and naive masking techniques.

\section{Related Work}
Constrained decoding in neural text generation has been an active area of research, particularly for controlling output quality and ensuring adherence to specific constraints. Traditional approaches, such as setting logits for undesirable tokens to $-\infty$, have proven effective at strictly excluding certain tokens, yet they often disrupt the underlying probability distribution of the model \cite{welleck2019neuraltextgenerationunlikelihood}.

In neural machine translation, \cite{susanto2020lexicallyconstrainedneuralmachine} explore lexically constrained decoding techniques using the Levenshtein Transformer, demonstrating that dynamic programming can enforce constraints, though at the expense of distorting the native conditional distribution. Recent advances have aimed to preserve the original distribution more faithfully while enforcing constraints. For example, \cite{park2024grammaraligneddecoding} propose Grammar-Aligned Decoding that integrates syntactic constraints into the decoding process, while \cite{chewi2021rejectionsamplingshapeconstraineddistributions} present a rejection sampling method tailored for shape-constrained distributions. Furthermore, \cite{melcer2024approximatelyaligneddecoding} introduce Approximately Aligned Decoding to reduce distributional shifts through refined re-normalization, and \cite{lew2023sequentialmontecarlosteering} explore Sequential Monte Carlo methods for constrained decoding. 

Our work builds on these approaches by formulating constrained decoding as a KL divergence minimization problem, resulting in a logit processor that preserves the model’s inherent conditional probabilities while excluding banned tokens.

\section{Methodology}
In autoregressive language modeling, the model generates the next token $x_t$ based on the conditional probability distribution:
\begin{equation}
    P(x_t \mid x_{<t}),
\end{equation}
where $x_{<t} = (x_1, x_2, \dots, x_{t-1})$ denotes the sequence of previously generated tokens. Let $\Omega$ denote the full vocabulary and $B \subset \Omega$ be the set of banned tokens (e.g., tokens corresponding to certain languages). Our goal is to enforce the constraint that no token from $B$ is generated while preserving the original probability distribution as faithfully as possible.

\subsection{Conditional Distribution under Constraints}
We define the desired constrained distribution $Q(x_t)$ as the conditional probability given that $x_t \notin B$:
\begin{equation}
    Q(x_t) = P(x_t \mid x_{<t},\, x_t \notin B).
\end{equation}
By the definition of conditional probability, we have:
\begin{equation}
    Q(x_t) =
    \begin{cases}
      \displaystyle \frac{P(x_t \mid x_{<t})}{\sum_{j \notin B} P(j \mid x_{<t})}, & \text{if } x_t \notin B, \\[1mm]
      0, & \text{if } x_t \in B.
    \end{cases}
    \label{eq:conditional}
\end{equation}
This formulation ensures that $\sum_{x_t \notin B} Q(x_t) = 1$, thereby preserving the total probability mass over the allowed tokens.

\subsection{Hypothesis on Distributional Distortion with Naive Masking}
We hypothesize that a naive logit masking approach, which sets the logits of banned tokens to $-\infty$, induces a re-normalized distribution that distorts the original probability mass of allowed tokens. Formally, let the raw logits be $\mathbf{z}$ such that
\[
P(x_t \mid x_{<t}) = \mathrm{softmax}(\mathbf{z}).
\]
A naive masking procedure defines a modified distribution $Q'(x_t)$ as:
\begin{equation}
    Q'(x_t) =
    \begin{cases}
      \displaystyle \frac{P(x_t \mid x_{<t})}{Z}, & \text{if } x_t \notin B, \\
      0, & \text{if } x_t \in B,
    \end{cases}
\end{equation}
with normalization constant
\[
Z = \sum_{j \notin B} P(j \mid x_{<t}).
\]
If a significant probability mass is assigned to tokens in $B$, then $Z \ll 1$, and the scaling factor $1/Z$ amplifies the differences among allowed tokens. This can lead to increased variance in the output quality, as small fluctuations in the original distribution are exaggerated after re-normalization. In contrast, our (G)I-DLE approach minimizes the KL divergence between the desired constrained distribution and the original distribution, thereby reducing such distortions.

\subsection{KL Divergence Minimization Framework}
A principled approach to obtain $Q(x_t)$ is to solve the following optimization problem:
\begin{align}
    \min_{Q}\; & D_{\mathrm{KL}}(Q \parallel P) \nonumber \\
    \text{subject to } & Q(x_t)=0 \quad \forall\, x_t \in B, \label{eq:kl_opt} \\
    & \sum_{x_t \notin B} Q(x_t)=1, \nonumber
\end{align}
where the KL divergence is defined as:
\begin{equation}
    D_{\mathrm{KL}}(Q \parallel P) = \sum_{x_t \in \Omega} Q(x_t) \log \frac{Q(x_t)}{P(x_t \mid x_{<t})}.
\end{equation}
It can be shown that the optimal solution to \eqref{eq:kl_opt} is given by Equation~\eqref{eq:conditional}. This demonstrates that our constrained distribution $Q(x_t)$ minimizes the divergence from the original distribution while strictly enforcing the exclusion of banned tokens.

\subsection{Logit Processor Implementation}
In practice, language models output \emph{logits} $\mathbf{z}$, with:
\[
P(x_t \mid x_{<t}) = \mathrm{softmax}(\mathbf{z}).
\]
To enforce the constraint, we adjust the logits for banned tokens by setting them to $-\infty$ and re-normalize the remaining logits. For each token index $i$:
\begin{equation}
    \tilde{z}_i =
    \begin{cases}
      \log P(i \mid x_{<t}) - \log\left(\sum_{j \notin B} P(j \mid x_{<t})\right), & i \notin B, \\
      -\infty, & i \in B.
    \end{cases}
\end{equation}
This transformation ensures that applying softmax to $\tilde{\mathbf{z}}$ yields the distribution $Q(x_t)$, thereby preserving the relative probabilities among allowed tokens.

\subsection{Theoretical and Practical Considerations}
\begin{itemize}
    \item \textbf{Distribution Preservation:} Minimizing the KL divergence ensures that the allowed token distribution remains as close as possible to the original model prediction.
    \item \textbf{Computational Efficiency:} The re-normalization step, while incurring additional computation over allowed tokens, remains efficient in practice, particularly when the banned set $B$ is small.
    \item \textbf{Generalizability:} Our framework can be extended to enforce multiple types of constraints, including stylistic, syntactic, or domain-specific restrictions.
    \item \textbf{Seamless Integration:} The proposed logit processor can be integrated into existing decoding pipelines without modifying the underlying model architecture.
\end{itemize}

\section{Experiments}

In this section, we evaluate our proposed (G)I-DLE method on the K\(^2\)-Eval dataset \cite{son2024llmasajudgerewardmodel}. K\(^2\)-Eval comprises 90 open-ended questions designed to assess the fluency, logical reasoning, and cultural appropriateness of responses in Korean. Unlike many benchmarks tailored for Western-centric domains, K\(^2\)-Eval focuses on Korea-related knowledge, making it particularly suitable for evaluating multilingual models like Qwen2.5 \cite{qwen2025qwen25technicalreport} in culturally specialized settings.

\subsection{Experimental Setup}
Due to resource constraints, our experiments include Qwen2.5 models with sizes ranging from 1.5B to 14B. We compare three generation methods:
\begin{enumerate}
    \item \textbf{Baseline Generation (Experiment 0):} Standard autoregressive generation without any logit processing.
    \item \textbf{Masked Generation (Experiment 1):} Generation where tokens corresponding to Chinese (as well as Japanese and Russian) characters are naively blocked by setting their logits to $-\infty$. Although this approach excludes the banned tokens, it can significantly distort the underlying distribution—particularly in a multilingual model where banned tokens may carry a substantial probability mass.
    \item \textbf{(G)I-DLE (Experiment 2):} Our proposed method, which leverages KL divergence minimization to preserve the model's conditional probability distribution while excluding banned tokens. This approach mitigates the distortion introduced by harsh masking.
\end{enumerate}

Generated responses are evaluated using an LLM-as-a-Judge framework (based on GPT-4o), focusing on four key metrics:
\begin{itemize}
    \item \textbf{Overall Score:} A composite measure reflecting fluency, logical consistency, and cultural appropriateness.
    \item \textbf{Fluency:} The naturalness and coherence of the language.
    \item \textbf{Logic:} The clarity and consistency of reasoning.
    \item \textbf{Cultural Appropriateness:} The alignment of responses with Korea-specific linguistic and cultural norms.
\end{itemize}

\subsection{Results}
Table~\ref{tab:results} summarizes the evaluation results across the different Qwen2.5 model sizes and experimental settings. The \textit{mean} column denotes the average evaluation score (on a scale from 1 to 5), and the \textit{var} column reports the variance of the scores.

\begin{table}[ht]
\centering
\caption{Evaluation Results on the K\(^2\)-Eval Dataset (Models Sorted in Ascending Order). Experiment settings: 0 -- Baseline, 1 -- Masked Generation, 2 -- (G)I-DLE.}
\label{tab:results}
\begin{tabular}{lccc}
\hline
\textbf{Model} & \textbf{Experiment} & \textbf{Mean Score} & \textbf{Variance} \\
\hline
Qwen2.5-1.5B-Instruct & 0 & 3.266667 & 1.793258 \\
Qwen2.5-1.5B-Instruct & 1 & 3.388889 & 1.858302 \\
Qwen2.5-1.5B-Instruct & 2 & 3.455556 & 1.981149 \\
Qwen2.5-3B-Instruct   & 0 & 4.483333 & 0.856461 \\
Qwen2.5-3B-Instruct   & 1 & 4.516667 & 0.822753 \\
Qwen2.5-3B-Instruct   & 2 & 4.555556 & 0.833958 \\
Qwen2.5-7B-Instruct   & 0 & 4.716667 & 0.517135 \\
Qwen2.5-7B-Instruct   & 1 & 4.900000 & 0.135955 \\
Qwen2.5-7B-Instruct   & 2 & 4.900000 & 0.113483 \\
Qwen2.5-14B-Instruct  & 0 & 4.911111 & 0.216729 \\
Qwen2.5-14B-Instruct  & 1 & 4.933333 & 0.085393 \\
Qwen2.5-14B-Instruct  & 2 & 4.966667 & 0.055056 \\
\hline
\end{tabular}
\end{table}

\subsection{Discussion of Experimental Results}
Our experimental results reveal several key insights:
\begin{enumerate}
    \item \textbf{Performance Scaling with Model Size:} Larger models (e.g., Qwen2.5-14B-Instruct) achieve substantially higher mean scores (approximately 4.91 to 4.97) compared to smaller models (Qwen2.5-1.5B-Instruct with scores around 3.27 to 3.46). This trend underscores the benefits of increased model capacity in generating high-quality, culturally appropriate responses.
    
    \item \textbf{Effectiveness of (G)I-DLE:} Across all model sizes, the (G)I-DLE method (Experiment 2) consistently yields the highest mean scores, accompanied by a significant reduction in variance. For instance, Qwen2.5-14B-Instruct shows a decrease in variance from 0.216729 (baseline) to 0.055056 under (G)I-DLE. This indicates that our method not only enhances average performance but also stabilizes output quality.
    
    \item \textbf{Limitations of Naive Masking:} The naive masked generation approach (Experiment 1) produces only marginal improvements in mean score over the baseline while exhibiting higher variance, particularly for smaller models. This finding supports our hypothesis that harsh logit suppression distorts the posterior distribution—especially when the banned tokens constitute a significant portion of the model’s probability mass—and leads to increased variability in output quality.
    
    \item \textbf{Domain-Specific Considerations:} Although Qwen2.5 is a multilingual model, the K\(^2\)-Eval dataset focuses on evaluating Korean language fluency and cultural appropriateness. Naively masking tokens from other languages may inadvertently disrupt the generation process, as the model’s training data allocates a non-negligible probability mass to these tokens. Our (G)I-DLE approach mitigates this issue by re-normalizing allowed token probabilities, thereby preserving the output quality in the target domain.
\end{enumerate}

Overall, our results validate the efficacy of (G)I-DLE in mitigating the adverse effects of naive token masking. The experimental evidence confirms that while naive masking enforces token exclusion, it introduces significant distributional distortions that increase output variance. In contrast, (G)I-DLE preserves the original conditional distribution, yielding outputs that are more fluent, logically coherent, and culturally appropriate.

\section{Discussion}
In this work, we introduced (G)I-DLE, a novel logit processing method for constrained decoding that leverages KL divergence minimization to preserve the conditional probability distribution of autoregressive language models while excluding banned tokens. Our theoretical analysis hypothesized that naive logit masking—by setting banned token logits to $-\infty$—forces a re-normalization that amplifies relative differences among allowed tokens, thereby increasing output variance. Our experimental results on the K\(^2\)-Eval dataset corroborate this hypothesis, as (G)I-DLE consistently achieves higher mean evaluation scores with substantially reduced variance compared to both baseline generation and naive masking approaches.

These findings underscore the importance of preserving the original probability distribution during constrained decoding, particularly in domain-specific applications such as culturally appropriate Korean text generation. Future work will extend this framework to multi-turn dialogues and explore additional constraint types to further enhance generation quality.

\section{Conclusion}
We have presented (G)I-DLE, a KL divergence minimization-based logit processor for constrained decoding. By preserving the inherent conditional probability distribution of autoregressive language models while excluding undesirable tokens, (G)I-DLE overcomes the limitations of naive masking techniques that introduce significant distributional distortions. Experiments on the K\(^2\)-Eval dataset—designed to assess Korean language fluency, logical reasoning, and cultural appropriateness—demonstrate that (G)I-DLE produces more fluent, logically coherent, and stable outputs across Qwen2.5 models ranging from 1.5B to 14B. Our results highlight the potential of distribution-preserving methods in constrained decoding and pave the way for further research into more complex, domain-specific constraint handling.

\appendix
\section{LLM-as-a-Judge Configuration}
For evaluation, we employed an LLM-as-a-Judge framework using GPT-4o. The evaluation prompt template (JUDGE\_TEMPLATE) used for single-turn evaluation is as follows:

\begin{verbatim}
JUDGE_TEMPLATE = {
    "single_turn": """You are tasked with evaluating the responses 
generated by a Korean language model. 
Please evaluate the response according to the following criteria:
1. Completeness: Does the response fully address the question?
2. Logical Consistency: Is the reasoning clear and logically sound?
3. Fluency: Is the language natural and coherent?
4. Cultural Appropriateness: Does the response adhere 
to Korean linguistic and cultural norms?
Provide your evaluation in 4-5 sentences, 
then assign a score from 1 to 5 in the following format:
Evaluation: [Your evaluation here]
Score: [Numeric score here]"""}
\end{verbatim}

\clearpage
\bibliographystyle{plain}
\bibliography{references}

\end{document}